\documentclass[11pt]{article}
\usepackage{times}
\usepackage{geometry}
\usepackage[utf8]{inputenc}
\usepackage{enumitem,amssymb}
\usepackage{ragged2e}
\usepackage{graphicx}
\usepackage{comment}
\usepackage{multicol}
\usepackage[usenames]{xcolor} 
\definecolor{xlinkcolor}{cmyk}{1,1,0,0}
\usepackage{url}
\usepackage[
 colorlinks=true,    
 linkcolor=xlinkcolor,     
 citecolor=xlinkcolor,     
 filecolor=xlinkcolor,  
 urlcolor=xlinkcolor,      
 final=true
]{hyperref}

\usepackage[super,sort&compress]{natbib}
\usepackage{enumitem}
\setenumerate{itemsep=0mm}

\begin{document}
\begin{raggedright} 
\huge
Snowmass2021 - Letter of Interest \hfill \\[+1em]
\textit{\textbf{The GRAMS Project: \\\Large MeV Gamma-Ray Observations and Antimatter-Based Dark Matter Searches}} \hfill \\[+1em]
\end{raggedright}

\normalsize

\noindent {\large \bf Thematic Areas:}  (check all that apply $\square$/$\blacksquare$)\\
\noindent $\blacksquare$ (CF1) Dark Matter: Particle Like \\
\noindent $\square$ (CF2) Dark Matter: Wavelike  \\ 
\noindent $\blacksquare$ (CF3) Dark Matter: Cosmic Probes  \\
\noindent $\square$ (CF4) Dark Energy and Cosmic Acceleration: The Modern Universe \\
\noindent $\square$ (CF5) Dark Energy and Cosmic Acceleration: Cosmic Dawn and Before \\
\noindent $\square$ (CF6) Dark Energy and Cosmic Acceleration: Complementarity of Probes and New Facilities \\
\noindent $\square$ (CF7) Cosmic Probes of Fundamental Physics \\
\noindent $\square$ (Other) {\it [Please specify frontier/topical group]} \\

\noindent {\large \bf Contact Information:} \\ 
Submitter Name/Institution: Georgia Karagiorgi/Columbia University \\
Collaboration: GRAMS (Gamma-Ray and AntiMatter Survey) \\
Contact Email: georgia@nevis.columbia.edu \\

\noindent {\large \bf Authors:} \\
Tsuguo Aramaki$^{1,2}$, Jonathan Asaadi$^{3}$, Yuto Ichinohe$^{4}$, Yoshiyuki Inoue$^{5}$, Georgia Karagiorgi$^6$, Jon Leyva$^2$, Reshmi Mukherjee$^6$, Hirokazu Odaka$^8$, Kerstin Perez$^9$, William Seligman$^6$, Satoshi Takashima$^8$, Naomi Tsuji$^5$, Hiroki Yoneda$^5$ \vspace{0.3cm}\\
\noindent\textit{$^{1}$Northeastern University, USA\\
$^{2}$SLAC National Accelerator Laboratory, USA\\
$^{3}$University of Texas at Arlington, USA\\
$^{4}$Rikkyo University, Japan\\
$^{5}$RIKEN, Japan\\
$^{6}$Columbia University, USA\\
$^{7}$Barnard College, Columbia University, USA\\
$^{8}$University of Tokyo, Japan\\
$^{9}$Massachusetts Institute of Technology, USA\\}

\noindent {\large \bf Abstract:} \\
The Gamma-Ray and AntiMatter Survey (GRAMS) project is a proposed next-generation balloon/satellite mission targeting both MeV gamma-ray observations and antimatter-based dark matter searches. A cost-effective, large-scale Liquid Argon Time Projection Chamber (LArTPC) detector technology will allow GRAMS to have a significantly improved sensitivity to MeV gamma rays while extensively probing dark matter parameter space via antimatter measurements. 
\clearpage

\noindent {\large \bf Project Overview:} 

 The Gamma-Ray and AntiMatter Survey (GRAMS) is a novel telescope that uses advanced technology to simultaneously target both
astrophysical observations with MeV gamma rays and an indirect dark matter search with
antimatter \cite{Aramaki_2020}.
The project targets measurements in the MeV-scale energy range, where astrophysical observations are sparse, and where new measurements would be valuable to the astrophysics and astronomy communities \cite{Takahashi_2013}. 
The GRAMS detector configuration also offers sensitivity to antiparticles from dark matter annihilation or decay. In particular, low-energy antideuterons can provide essentially background-free dark matter signatures while uniquely exploring dark matter parameter space \cite{Donato_2000,Baer_2005,Donato_2008,Ibarra_2013,Aramaki_2016a}.

GRAMS will make use of a novel detector technology---that of a liquid argon time projection chamber (LArTPC) \cite{Rubbia_1977}. The GRAMS detector is cost-effective, considering that argon is both naturally abundant and low-cost, which allows a large-scale detector to be deployed, unlike previous and current experiments with semiconductor or scintillation detectors. A LArTPC detector provides three-dimensional particle tracking capability by measuring ionization charge and scintillation light produced by particles entering or created in the argon medium. This technology is currently in use for and under continued development by the neutrino and direct dark matter search experimental communities (see, e.g.~\cite{Acciarri_2017,Antonello_2015,Abi_2017,Abi_2020,Alton2010}, for detectors deployed at sea level or deep underground). The proposed research will take advantage of recent developments in detector instrumentation to enable a LArTPC-based Compton telescope concept for high-altitude air balloon or satellite deployment, and with sensitivity to both low-energy and antiparticle signals. An incoming antiparticle stopped inside the detector can form an exotic atom with an Argon nucleus. Incoming antiparticles may be identified through the measurements of atomic x-rays and annihilation products (such as pions and protons) produced in the decay of the exotic atom  \cite{Aramaki_2020,Mori_2002,Aramaki_2013,Aramaki_2016b}.


The LArTPC technology provides unique advantages over conventional Compton telescope technology, which traditionally has made use of solid or gas detectors. Specifically, (i) gas detectors are limited in effective area, due to their low density, while (ii) detectors made of solid materials such as semiconductors (Si, CdTe or Ge), which are often used because of their high density, are limited in active detector volume, and thus effective area, due to necessary embedded supporting materials and readout electronics, as well as due to scalability cost constraints. Liquid argon, being relatively inexpensive and easily sourced, evades density as well as scalability limitations. Moreover, the LArTPC technology provides a high detection efficiency for antimatter detection, since there is almost no dead material inside the detector, unlike semiconductor detectors with mounting frames and pre-amps near-by.

\noindent {\large \bf Science Motivation:} 

In recent astrophysical observations, the AGILE \cite{Tavani_2009}, Fermi \cite{Abdollahi_2020}, INTEGRAL \cite{Knodlseder_2005}, NuSTAR \cite{Harrison_2013}, and Swift \cite{Oh_2018} satellite instruments opened new windows to observe astrophysical phenomena in the energy domains of hard X-rays ($\lesssim200$~keV) and high-energy gamma-rays (above 100~MeV), respectively. 
However, gamma rays in the MeV energy range have not yet been well-explored. COMPTEL \cite{Schoenfelder_2000} onboard the CGRO satellite launched in 1991 has produced the first catalogue of MeV gamma-ray sources, but only approximately 30 objects have been detected thus far. Since COMPTEL and INTEGRAL, scientific progress in the 0.1 to 100~MeV range, commonly known as the ``MeV gap,'' has been  somewhat limited. 

Through MeV searches, GRAMS will enable the study of energetic particle acceleration, since the transition from thermal to non-thermal physical processes occurs in the MeV energy region. These phenomena can be seen in various celestial objects, e.g., reconnection in the Sun, magnetosphere of neutron stars, and relativistic outflows generated by black holes, gamma-ray bursts, and so on \cite{Lin_2003,Zdziarski_2017,IceCube_2018, Inoue_2019, Meszaros_2013}. Those class of objects are known to have spectral features in the MeV gamma-ray band. GRAMS could also detect MeV gamma rays associated with neutron star mergers and gamma-ray bursts \cite{Abbott_2017}, lending itself to Gravitational Wave Electromagnetic Counterpart (GW-EM) astrophysics.

GRAMS also enables us to investigate the evolutionary history of the Universe. The measurement of the cosmic MeV gamma-ray background spectrum and anisotropy tell the entire history of MeV gamma-ray activity in the Universe \cite{Inoue_2013}. Furthermore, some of blazars are known to be found even at very high redshift $z\gtrsim 5$ \cite{Sbarrato_2013}, because they have a spectral peak in the MeV energy band, so-called MeV blazars. Through observations of those MeV blazars, GRAMS can investigate the growth history of supermassive black hole and their activity history in the early Universe.

The MeV range is the domain of nuclear gamma-ray lines and is perhaps the only part of the electromagnetic spectrum where it is possible to directly observe nuclear processes. With sufficient energy resolution, GRAMS can offer unique opportunity to directly probe nucleosynthesis processes in various astrophysical environment such as in the Galactic Center \cite{Knodlseder_2005} and in accretion flows \cite{Kafexhiu_2019}. 
The 511 keV positron annihilation emission from the Galactic Center region is still a mystery and identifying the source of positrons is of scientific interest for the understanding of stellar explosions as well as the study of compact objects. GRAMS will be sensitive to the 1.8 MeV gamma-ray emission line, produced by the radioactive decay of $^{26}$Al, which can be used to trace regions with massive young stars throughout the Milky Way. The energy range of GRAMS will cover the regime of the characteristic pion bump which may be produced due to hadronic emission processes in supernova remnants.
Gamma-ray line measurements are also important for multi-messenger astronomy, studying transient phenomena. For example, these measurements can directly probe the r-process nucleosynthesis in neutron star mergers associated with gravitational waves \cite{Hotokezaka_2016}, and can determine the explosion mechanism of type-Ia supernovae--the cosmic standard candles \cite{Horiuchi_2010,Summa_2013}. While this is challenging, the study of gamma rays from galactic binary neutron star merger remnants may prove to be possible with a next-generation instrument such as GRAMS, and can also help unresolved issues in nuclear astrophysics. A potential detection of diffuse flux from these objects could help trace the galactic spatial distribution of these objects over long timescales \cite{Wu_2019}. 



Finally, GRAMS MeV gamma-ray and antimatter measurements will provide unique opportunities to search for dark matter candidates. In particular, the gamma-ray observations in the ``MeV-gap'' can provide strong constraints on light dark matter models in the MeV mass range as well as ultralight primordial black holes that can comprise all or part of dark matter \cite{Essig_2013,Laha_2020,Carr_2020}. On the other hand, antimatter measurements can search for ``conventional'' dark matter models in the GeV and higher mass range, such as weakly interacting massive particles \cite{Donato_2000,Baer_2005,Donato_2008,Dal_2014,Randall_2020}. Low-energy antiparticles, especially antideuterons, can be background-free dark matter signatures, which will allow GRAMS to investigate the possible dark matter signals suggested in the Fermi gamma-ray observations and AMS-02 antiproton measurements \cite{Aramaki_2020,Calore_2015,Daylan_2016,Abazajian_2016,Cui_2017,Ackermann_2017,Alvarez_2020}.



\noindent {\large \bf Summary:}

A next-generation instrument such as GRAMS could potentially be transformative in the ``medium energy'' gamma-ray astrophysics regime, offering a sensitive look in a band that has been long-unexplored. A sensitive instrument in the 0.1-100 MeV band is needed that will have broad spectral coverage, wide field of view, improved angular resolution, sensitivity to gamma-ray line emission even better than COMPTEL, and polarization sensitivity that will enable it to measure polarization fraction in GRBs, pulsars and active galaxies with SMBHs. The themes of scientific exploration with GRAMS will encompass radioactivity and antimatter, cosmic-ray physics, black holes, neutron stars and pulsars, and fundamental physics topics, including dark matter annihilation and decay. The GRAMS collaboration welcomes new members, and plans to submit a ``Whitepaper'' to the Snowmass 2021 community planning process.

\clearpage

\bibliographystyle{unsrt}
\bibliography{main}{}

\begin{thebibliography}{10}

\bibitem{Aramaki_2020}
T.~Aramaki, P.H. Adrian, G.~Karagiorgi, and H.~Odaka.
\newblock Dual mev gamma-ray and dark matter observatory - grams project.
\newblock {\em Astroparticle Physics}, 114:107–114, Jan 2020.

\bibitem{Takahashi_2013}
T.~Takahashi, Y.~Uchiyama, and Ł. Stawarz.
\newblock Multiwavelength astronomy and cta: X-rays.
\newblock {\em Astroparticle Physics}, 43:142–154, Mar 2013.

\bibitem{Donato_2000}
F.~Donato, N.~Fornengo, and P.~Salati.
\newblock Antideuterons as a signature of supersymmetric dark matter.
\newblock {\em Physical Review D}, 62(4), Jul 2000.

\bibitem{Baer_2005}
Howard Baer and Stefano Profumo.
\newblock Low energy antideuterons: shedding light on dark matter.
\newblock {\em Journal of Cosmology and Astroparticle Physics},
  2005(12):008–008, Dec 2005.

\bibitem{Donato_2008}
F.~Donato, N.~Fornengo, and D.~Maurin.
\newblock Antideuteron fluxes from dark matter annihilation in diffusion
  models.
\newblock {\em Physical Review D}, 78(4), Aug 2008.

\bibitem{Ibarra_2013}
A.~Ibarra and S.~Wild.
\newblock Determination of the cosmic antideuteron flux in a monte carlo
  approach.
\newblock {\em Physical Review D}, 88(2), Jul 2013.

\bibitem{Aramaki_2016a}
T.~Aramaki, S.~Boggs, S.~Bufalino, L.~Dal, P.~von Doetinchem, F.~Donato,
  N.~Fornengo, H.~Fuke, M.~Grefe, C.J. Hailey, and et~al.
\newblock Review of the theoretical and experimental status of dark matter
  identification with cosmic-ray antideuterons.
\newblock {\em Physics Reports}, 618:1–37, Mar 2016.

\bibitem{Rubbia_1977}
C.~Rubbia.
\newblock The liquid-argon time projection chamber: a new concept for neutrino
  detectors.
\newblock Technical report, 1977.

\bibitem{Acciarri_2017}
R.~Acciarri, C.~Adams, R.~An, A.~Aparicio, S.~Aponte, J.~Asaadi, M.~Auger,
  N.~Ayoub, L.~Bagby, B.~Baller, and et~al.
\newblock Design and construction of the microboone detector.
\newblock {\em Journal of Instrumentation}, 12(02):P02017–P02017, Feb 2017.

\bibitem{Antonello_2015}
M.~Antonello, Z.~Djurcic, V.~Genty, J.~Zennamo, A.~Dermenev, G.H. Collin,
  B.~Russell, M.~Richardson, A.~Marchionni, E.~Church, et~al.
\newblock A proposal for a three detector short-baseline neutrino oscillation
  program in the fermilab booster neutrino beam.
\newblock Technical report, 2015.

\bibitem{Abi_2017}
B.~Abi, R.~Acciarri, M.A. Acero, M.~Adamowski, C.~Adams, D.L. Adams,
  P.~Adamson, M.~Adinolfi, Z.~Ahmad, C.H. Albright, et~al.
\newblock The single-phase protodune technical design report.
\newblock {\em arXiv preprint arXiv:1706.07081}, 2017.

\bibitem{Abi_2020}
B.~Abi, R~Acciarri, M.A. Acero, G.~Adamov, D.~Adams, M.~Adinolfi, Z.~Ahmad,
  J.~Ahmed, T.~Alion, S.A. Monsalve, et~al.
\newblock Deep underground neutrino experiment (dune), far detector technical
  design report, volume iv far detector single-phase technology.
\newblock {\em arXiv preprint arXiv:2002.03010}, 2020.

\bibitem{Alton2010}
D.~Alton, D.~Durben, M.~Keeter, K.~Zehfus, et~al.
\newblock Darkside-50: A direct search for dark matter with new techniques for
  reducing background.
\newblock {\em DarkSide50 DOE Project Narrative FNAL. pdf}, 2010.

\bibitem{Mori_2002}
K.~Mori, C.J. Hailey, E.A. Baltz, W.W. Craig, M.~Kamionkowski, W.T. Serber, and
  P.~Ullio.
\newblock A novel antimatter detector based on x‐ray deexcitation of exotic
  atoms.
\newblock {\em The Astrophysical Journal}, 566(1):604–616, Feb 2002.

\bibitem{Aramaki_2013}
T.~Aramaki, S.K. Chan, W.W. Craig, L.~Fabris, F.~Gahbauer, C.J. Hailey, J.E.
  Koglin, N.~Madden, K.~Mori, H.T. Yu, and et~al.
\newblock A measurement of atomic x-ray yields in exotic atoms and implications
  for an antideuteron-based dark matter search.
\newblock {\em Astroparticle Physics}, 49:52–62, Sep 2013.

\bibitem{Aramaki_2016b}
T.~Aramaki, C.J. Hailey, S.E. Boggs, P.~von Doetinchem, H.~Fuke, S.I. Mognet,
  R.A. Ong, K.~Perez, and J.~Zweerink.
\newblock Antideuteron sensitivity for the gaps experiment.
\newblock {\em Astroparticle Physics}, 74:6–13, Feb 2016.

\bibitem{Tavani_2009}
M.~Tavani, G.~Barbiellini, A.~Argan, F.~Boffelli, A.~Bulgarelli, P.~Caraveo,
  P.~W. Cattaneo, A.~W. Chen, V.~Cocco, E.~Costa, and et~al.
\newblock The agile mission.
\newblock {\em Astronomy \& Astrophysics}, 502(3):995–1013, Jan 2009.

\bibitem{Abdollahi_2020}
S.~Abdollahi, F.~Acero, M.~Ackermann, M.~Ajello, W.~B. Atwood, M.~Axelsson,
  L.~Baldini, J.~Ballet, G.~Barbiellini, D.~Bastieri, and et~al.
\newblock Fermi large area telescope fourth source catalog.
\newblock {\em The Astrophysical Journal Supplement Series}, 247(1):33, Mar
  2020.

\bibitem{Knodlseder_2005}
J.~Knödlseder, P.~Jean, V.~Lonjou, G.~Weidenspointner, N.~Guessoum,
  W.~Gillard, G.~Skinner, P.~von Ballmoos, G.~Vedrenne, J.-P. Roques, and
  et~al.
\newblock The all-sky distribution of 511 kev electron-positron annihilation
  emission.
\newblock {\em Astronomy \& Astrophysics}, 441(2):513–532, Sep 2005.

\bibitem{Harrison_2013}
F.A. Harrison, W.W. Craig, F.E. Christensen, C.J. Hailey, W.W. Zhang, S.E.
  Boggs, D.~Stern, W.R. Cook, K.~Forster, P.~Giommi, and et~al.
\newblock The nuclear spectroscopic telescope array(nustar) high-energy x-ray
  mission.
\newblock {\em The Astrophysical Journal}, 770(2):103, May 2013.

\bibitem{Oh_2018}
K.~Oh, M.l Koss, C.B. Markwardt, K.~Schawinski, W.H. Baumgartner, S.D.
  Barthelmy, S.B. Cenko, N.~Gehrels, R.~Mushotzky, A.~Petulante, and et~al.
\newblock The 105-month swift-bat all-sky hard x-ray survey.
\newblock {\em The Astrophysical Journal Supplement Series}, 235(1):4, May
  2018.

\bibitem{Schoenfelder_2000}
V.~Schönfelder, K.~Bennett, J.~J. Blom, H.~Bloemen, W.~Collmar, A.~Connors,
  R.~Diehl, W.~Hermsen, A.~Iyudin, R.~M. Kippen, and et~al.
\newblock The first comptel source catalogue.
\newblock {\em Astronomy and Astrophysics Supplement Series}, 143(2):145–179,
  Apr 2000.

\bibitem{Lin_2003}
R.P.~c Lin, S~Krucker, G.J. Hurford, D.M. Smith, H.S. Hudson, G.D. Holman, R.A.
  Schwartz, B.R. Dennis, G.H. Share, R.J. Murphy, et~al.
\newblock Rhessi observations of particle acceleration and energy release in an
  intense solar gamma-ray line flare.
\newblock {\em The Astrophysical Journal Letters}, 595(2):L69, 2003.

\bibitem{Zdziarski_2017}
A.A. Zdziarski, D.~Malyshev, M.~Chernyakova, and G.G. Pooley.
\newblock High-energy gamma-rays from cyg x-1.
\newblock {\em Monthly Notices of the Royal Astronomical Society},
  471(3):3657–3667, Jul 2017.

\bibitem{IceCube_2018}
IceCube Collaboration et~al.
\newblock Multimessenger observations of a flaring blazar coincident with
  high-energy neutrino icecube-170922a.
\newblock {\em Science}, 361(6398), 2018.

\bibitem{Inoue_2019}
Y.~Inoue, D.~Khangulyan, S.~Inoue, and A.~Doi.
\newblock On high-energy particles in accretion disk coronae of supermassive
  black holes: Implications for mev gamma-rays and high-energy neutrinos from
  agn cores.
\newblock {\em The Astrophysical Journal}, 880(1):40, Jul 2019.

\bibitem{Meszaros_2013}
Peter Mészáros.
\newblock Gamma ray bursts.
\newblock {\em Astroparticle Physics}, 43:134–141, Mar 2013.

\bibitem{Abbott_2017}
B.~P. Abbott, R.~Abbott, T.~D. Abbott, F.~Acernese, K.~Ackley, C.~Adams,
  T.~Adams, P.~Addesso, R.~X. Adhikari, V.~B. Adya, and et~al.
\newblock Multi-messenger observations of a binary neutron star merger.
\newblock {\em The Astrophysical Journal}, 848(2):L12, Oct 2017.

\bibitem{Inoue_2013}
Y.~Inoue, K.~Murase, G.M. Madejski, and Y.~Uchiyama.
\newblock Probing the cosmic x-ray and mev gamma-ray background radiation
  through the anisotropy.
\newblock {\em The Astrophysical Journal}, 776(1):33, Sep 2013.

\bibitem{Sbarrato_2013}
T.~Sbarrato, G.~Tagliaferri, G.~Ghisellini, M.~Perri, S.~Puccetti,
  M.~Baloković, M.~Nardini, D.~Stern, S.~E. Boggs, W.~N. Brandt, and et~al.
\newblock Nustardetection of the blazar b2 1023+25 at redshift 5.3.
\newblock {\em The Astrophysical Journal}, 777(2):147, Oct 2013.

\bibitem{Kafexhiu_2019}
E.~Kafexhiu, F.~Aharonian, and M.~Barkov.
\newblock Nuclear $\gamma$-ray emission from very hot accretion flows.
\newblock {\em Astronomy \& Astrophysics}, 623:A174, 2019.

\bibitem{Hotokezaka_2016}
K.~Hotokezaka, S.~Wanajo, M.~Tanaka, A.~Bamba, Y.~Terada, and T.~Piran.
\newblock Radioactive decay products in neutron star merger ejecta: heating
  efficiency and $\gamma$-ray emission.
\newblock {\em Monthly Notices of the Royal Astronomical Society},
  459(1):35–43, Mar 2016.

\bibitem{Horiuchi_2010}
S.~Horiuchi and J.F. Beacom.
\newblock Revealing type ia supernova physics with cosmic rates and nuclear
  gamma rays.
\newblock {\em The Astrophysical Journal}, 723(1):329–341, Oct 2010.

\bibitem{Summa_2013}
A.~Summa, A.~Ulyanov, M.~Kromer, S.~Boyer, F.~K. Röpke, S.~A. Sim, I.~R.
  Seitenzahl, M.~Fink, K.~Mannheim, R.~Pakmor, and et~al.
\newblock Gamma-ray diagnostics of type ia supernovae.
\newblock {\em Astronomy \& Astrophysics}, 554:A67, Jun 2013.

\bibitem{Wu_2019}
M.R. Wu, P.~Banerjee, B.D. Metzger, G.~Martínez-Pinedo, T.~Aramaki, E.~Burns,
  C.J. Hailey, J.~Barnes, and G.~Karagiorgi.
\newblock Finding the remnants of the milky way’s last neutron star mergers.
\newblock {\em The Astrophysical Journal}, 880(1):23, Jul 2019.

\bibitem{Essig_2013}
R.~Essig, E.~Kuflik, S.D. McDermott, T.~Volansky, and K.M. Zurek.
\newblock Constraining light dark matter with diffuse x-ray and gamma-ray
  observations.
\newblock {\em Journal of High Energy Physics}, 2013(11), Nov 2013.

\bibitem{Laha_2020}
Ra. Laha, J.B. Muñoz, and T.R. Slatyer.
\newblock Integral constraints on primordial black holes and particle dark
  matter.
\newblock {\em Physical Review D}, 101(12), Jun 2020.

\bibitem{Carr_2020}
B.~Carr, K.~Kohri, Y.~Sendouda, and J.~Yokoyama.
\newblock Constraints on primordial black holes.
\newblock {\em arXiv preprint arXiv:2002.12778}, 2020.

\bibitem{Dal_2014}
L.A. Dal and A.R. Raklev.
\newblock Antideuteron limits on decaying dark matter with a tuned formation
  model.
\newblock {\em Physical Review D}, 89(10), May 2014.

\bibitem{Randall_2020}
L.~Randall and W.L. Xu.
\newblock Searching for dark photon dark matter with cosmic ray antideuterons.
\newblock {\em Journal of High Energy Physics}, 2020(5), May 2020.

\bibitem{Calore_2015}
F.~Calore, I.~Cholis, C.~McCabe, and C.~Weniger.
\newblock A tale of tails: Dark matter interpretations of the fermi gev excess
  in light of background model systematics.
\newblock {\em Physical Review D}, 91(6), Mar 2015.

\bibitem{Daylan_2016}
T.~Daylan, D.P. Finkbeiner, D.~Hooper, T.~Linden, S.K.N. Portillo, N.L. Rodd,
  and T.R. Slatyer.
\newblock The characterization of the gamma-ray signal from the central milky
  way: A case for annihilating dark matter.
\newblock {\em Physics of the Dark Universe}, 12:1–23, Jun 2016.

\bibitem{Abazajian_2016}
K.N. Abazajian and R.E. Keeley.
\newblock Bright gamma-ray galactic center excess and dark dwarfs: Strong
  tension for dark matter annihilation despite milky way halo profile and
  diffuse emission uncertainties.
\newblock {\em Physical Review D}, 93(8), Apr 2016.

\bibitem{Cui_2017}
M.Y. Cui, Q.~Yuan, Y.L.S. Tsai, and Y.Z. Fan.
\newblock Possible dark matter annihilation signal in the ams-02 antiproton
  data.
\newblock {\em Physical Review Letters}, 118(19), May 2017.

\bibitem{Ackermann_2017}
M.~Ackermann, M.~Ajello, A.~Albert, W.~B. Atwood, L.~Baldini, J.~Ballet,
  G.~Barbiellini, D.~Bastieri, R.~Bellazzini, E.~Bissaldi, and et~al.
\newblock Thefermigalactic center gev excess and implications for dark matter.
\newblock {\em The Astrophysical Journal}, 840(1):43, May 2017.

\bibitem{Alvarez_2020}
A.~Alvarez, F.~Calore, A.~Genina, J.I. Read, P.D. Serpico, and B.~Zaldivar.
\newblock Dark matter constraints from dwarf galaxies with data-driven
  j-factors.
\newblock {\em arXiv preprint arXiv:2002.01229}, 2020.

\end{thebibliography}

\end{document}